# AI-Enabled Rapid Assembly of Thousands of Defect-Free Neutral Atom Arrays with Constant-time-overhead

Rui Lin[1,2,3], Han-Sen Zhong[4,*], You Li[1,2,3], Zhang-Rui Zhao[4], Le-Tian Zheng[1,2,3], Tai-Ran Hu[1,2,3], Hong-Ming Wu[1,2,3], Zhan Wu[1,2,3], Wei-Jie Ma[4], Yan Gao[4], Yi-Kang Zhu[5], Zhao-Feng Su[5], Wan-Li Ouyang[4], Yu-Chen Zhang[1,2,3], Jun Rui[1,2,3], Ming-Cheng Chen[1,2,3], Chao-Yang Lu[1,2,3,*], Jian-Wei Pan[1,2,3,*]

[1]Hefei National Research Center for Physical Sciences at the Microscale and School of Physical Sciences, University of Science and Technology of China, Hefei 230026, China

[2]Shanghai Research Center for Quantum Science and CAS Center for Excellence in Quantum Information and Quantum Physics, University of Science and Technology of China, Shanghai 201315, China

[3]Hefei National Laboratory, University of Science and Technology of China, Hefei 230088, China

[4]Shanghai Artificial Intelligence Laboratory, Shanghai 200232, China

[5]School of Computer Science and Technology and CAS Key Laboratory of Wireless-Optical Communications, University of Science and Technology of China, Hefei 230026, China

Emails: zhonghansen@pjlab.org.cn, cylu@ustc.edu.cn, pan@ustc.edu.cn

**Abstract**

Assembling increasingly larger-scale defect-free optical tweezer-trapped atom arrays is essential for quantum computation and quantum simulations based on atoms. Here, we propose an AI-enabled, rapid, constant-time-overhead rearrangement protocol, and we experimentally assemble defect-free 2D and 3D atom arrays with up to 2024 atoms with a constant time cost of 60 ms. The AI model calculates the holograms for real-time atom rearrangement. With precise controls over both position and phase, a high-speed spatial light modulator moves all the atoms simultaneously. This protocol can be readily used to generate defect-free arrays of tens of thousands of atoms with current technologies, and become a useful toolbox for quantum error correction.

**Main text**

Quantum computation and quantum simulation have the potential to solve certain problems that are intractable to their classical counterparts. Early evidence of quantum computational speed-up has been provided by recent noisy intermediate-scale quantum machines [1,2]. However, scaling up the number of qubits and reducing the gate error rate for large-scale fault-tolerant quantum computers still remain significant challenges.

Rydberg atom arrays have recently emerged as a promising platform for quantum computation and simulations, featuring excellent scalability [3], high gate fidelity [4–6], high parallelism, and all-to-all connectivity [7,8] in a flexible and dynamically reconfigurable architecture. With control of tens to hundreds of atoms, key ingredients in quantum error correction have been demonstrated, including coherent transport [7], individual addressing [8,9], mid-circuit readout [10–12], and logical operations [8]. In addition, strong Rydberg interactions and arbitrary geometrical configurations in this system have spurred significant experimental progress in quantum simulation, such as probing (2+1)D Ising quantum phase transitions [13,14], exploring quantum many-body scars [15], and preparation of quantum spin liquids [16].

To generate a defect-free atom array, many rearrangement protocols [17–24] have been proposed. The mainstream methods, as demonstrated in Refs. [17,18,20,22–24], exploited a set of movable tweezers generated by acousto-optic deflectors (AODs) to move single atoms one by one or row by row. However, this method is constrained by the linear time complexity with respect to the atom size, becoming unfeasible for large scales. Other approaches, reported in Refs. [19,21], attempted to move atoms globally, but were hindered by severe atom loss and slow computation speed, which have so far been limited to small-scale arrays with only a few dozen atoms.

In this work, we propose and implement an efficient AI-enabled protocol for rapid rearrangement of defect-free atom arrays at a constant time cost independent of the size. Our rearrangement protocol features high parallelism, empowered by the AI program that drives a high-speed spatial light modulator (SLM). The main idea is illustrated in Fig. 1. Starting from an initially randomly loaded 2D atom array, we first calculate the optimal matching path between the loaded atoms and target sites. The entire path is then divided into $N$ steps. For each small step, the AI model is used to calculate the hologram for the SLM and, in real time, the SLM simultaneously moves all atoms to their target positions. This method achieves a high level of parallelism and thus a constant-time performance.

Figure 2 shows a schematic of our experimental set-up. We use a high-speed SLM with 1-kHz refresh rate to imprint an appropriate hologram onto the incident laser beam, then focus the beam through a 0.55-NA objective to generate the optical tweezers. The holograms for the initial and user-defined arbitrary target arrays are calculated using the weighted Gerchberg-Saxton (WGS) algorithm [25,26] and further optimized for array uniformity [27]. The initial tweezer array is stochastically loaded with single $^{87}$Rb atoms from a magneto-optical trap with a loading probability of ~65%. After the loading, a fluorescence image is captured by an electron-multiplying charge-coupled device (EMCCD) to recognize the single atom occupancy on each site. We develop a 3-layer convolutional network to process the image, enabling a high imaging fidelity of 99.92% [27].

To transform the randomly loaded array to defect-free array, we first calculate the path matching, i.e., finding the optimal movement path for each atom, by minimizing the cost function $\Sigma d_{ij}^2$, where $d_{ij}$ represents the distance between the matching atom $i$ and target site $j$. The cost function $\Sigma d_{ij}^2$ is optimized to avoid path-collisions and minimize the longest movement path [19,27]. The matching is calculated using Hungarian algorithm [28], modified with a block decomposition scheme to enable parallel computation. This ensures a nearly constant computation time of approximately 5 ms for arrays ranging from 1,000 to 10,000 atoms [27].

Having obtained the optimal matching path, the next step is to move the atoms to their target positions. The central challenge is to precisely control the SLM refresh and minimize atom loss. We consider an optical tweezer initially at position $r_1$ with phase $\phi_1$, which, after the hologram switches, moves to position $r_2$ with phase $\phi_2$. During the SLM refresh, the light field approximately follows [27]:

$$E(r,t) = e^{-\frac{t}{T}} E_1(r) e^{i\phi_1} + \left(1 - e^{-\frac{t}{T}}\right) E_2(r) e^{i\phi_2},$$

where $T$ is the response time (~1 ms in our experiment), and $E_1(E_2)$ represents the initial (final) Gaussian field of the tweezer at $r_1(r_2)$. The dynamically transitioning field gradually moves an atom from $r_1$ to $r_2$.

As shown in Fig. S4, the key to a successful movement is to keep the step size $\Delta r$ and phase change $\Delta \phi$ sufficiently small; otherwise, substantial atom heating and loss would occur. Therefore, the whole movement path is divided into ~20 small steps. For each small step, we use linear interpolation to extract its position and phase [27].

The next task is to generate the corresponding hologram for each step. Traditional algorithms [25,26,29] for the hologram generation, typically based on iterative Fourier transform algorithm [30], however, leave the phase of each tweezer uncontrolled. In addition, the positional accuracy of these methods is relatively low, requiring an expansion of the computational space, which, however, comes with a price of significantly increased computational time.

To overcome these limitations, we develop an AI model to generate the hologram, which is capable of precise control over the position and phase of each tweezer, all while maintaining fast and constant-time computation. We design our AI model that calculates the hologram as a fully convolutional neural network (CNN). The architecture consists of three input convolutional layers, a sequence of three residual blocks and ends with an output convolution. Each layer is equipped with convolutional module for main feature extraction and rectified linear unit as the activation function. The residual blocks introduce skip connections for deeper extraction of high-dimensional details, and batch normalization is applied in every residual block to ensure consistent activation distributions across all layers. Containing complex frequency domain information, holograms are not suited for direct generation by a CNN, which excels at processing information in the planar position domain. Therefore, the network is designed to first generate the amplitude and phase in the position domain, and then apply an inverse fast Fourier transform (FFT) to transform these predictions into the final hologram in the frequency domain.

To train our AI model (see Fig. 3a), we simulate the coordinates of the tweezer array at each step during the rearrangement process and adopt the WGS algorithm to generate the corresponding holograms, thus constructing an on-the-fly dataset for model training. We expand the computational space for the hologram generation by 64 times (8 times in each dimension) to achieve the required positional accuracy. To adapt for the CNN, these holograms will be transformed into the amplitudes and phases in the position domain by FFT to extract tweezers phases and serve as the target supervision $\{A_{label}, \phi_{label}\}$. The network input amplitude is represented as a 2D interpolated weight image $A_{input}$, encoded by bilinearly interpolating the coordinates, while the input phase image $\phi_{input}$ is encoded by mapping the tweezers phases to the corresponding positions [27]. We optimize the model by minimizing the differences ($l_1$ norm for $A$ and $l_2$ norm for $\phi$) between the prediction $\{A_{output}, \phi_{output}\}$ and target supervision. In practical inference, the lightweight model can generate the accurate hologram in a short and constant computation time [27]. Figure 3b shows the precise control over the positions and phases of the tweezers achieved by the AI model.

Using the AI model, we calculate the hologram for each small step according to the extracted positions and phases of all optical tweezers. To speed up the computation, the holograms are calculated in parallel using two GPUs. Once a hologram is calculated, the SLM is refreshed immediately, enabling the next step of movement. In the first step, we turn off the unmatched traps, while the matched traps maintain their positions and phases. Next, all atoms are moved simultaneously, step by step, to their target positions. In the final step, with all the traps maintaining their positions and phases, we switch to the pre-defined target array, thus, a defect-free atom array with a high uniformity is achieved.

The dynamic atomic rearrangement process is illustrated in a Supplemental video, through the use of a Schrödinger's cat cartoon as an illustrative example, where each dot is the imaging of a single atom. We present the atom rearrangement results by highlighting key features of scale, efficiency, and time cost. Figs. 4a and 4b show a gallery of experimentally assembled defect-free 2D atom arrays with arbitrary configurations. Our largest 2D array contains 2024 atoms (out of 45×45=2025 sites), which, to our knowledge, is currently the largest defect-free atom array.

Our rearrangement protocol is also applied to construct 3D atom arrays. Here, the atom movement is restricted within the same layers to avoid atom loss when moving across layers, due to the weak axial confinement of the optical tweezers. We use the AI model to generate holograms for individual 2D layers and superpose corresponding Fresnel phases for the 3D rearrangement [27]. The assembly of atom arrays of trilayer cuboid and twisted graphene structures are shown in Fig. 4c and Fig. 4d, respectively. The typical filling fractions are 97.9(5)%.

Figure 5a shows the statistical distribution of the atom number in the target arrays from 1000 repetitions of the experiments. A single round of rearrangement results in a typical filling fraction of 99.0(2)%. The filling fraction is further improved to 99.6(2)% by performing a second round of rearrangement.

Figure 5b plots the time overhead for each subroutine for the atom rearrangement, including the path matching, the hologram computation on the two GPUs, and the SLM refresh, which costs 5 ms, 52 ms, and 20 ms, respectively. Remarkably, these time costs remain constant for different atom array size from 1,000 to 10,000. In the current work, as the GPU computation and SLM refresh are performed simultaneously, thus the entire rearrangement process takes less than 60 ms. The overall time cost of 60 ms can be further decreased by employing GPU clusters and upgrading SLMs with higher refresh rate. For example, new electro-optic SLMs, demonstrated in Refs. [31,32] with ~GHz refresh rate, can significantly speed up the rearrangement. It may also provide a fast, globally parallel method for the coherent transport and individual addressing of atomic qubits. Such an ability of global, arbitrary movement of atom arrays is also useful for quantum error correction.

In the near future, assembling defect-free arrays of tens of thousands of atoms is technologically feasible, by upgrading with higher-power lasers, higher-resolution SLM, objective with larger field of view, and vacuum system with longer lifetime [3,33]. Our work lays the groundwork for fault-tolerant quantum computation with large-scale atomic qubits. The large-scale atom arrays with arbitrary 2D or 3D configurations will also facilitate future explorations of novel quantum many-body phenomena.

**Figures**

**Fig. 1 | AI-enabled rearrangement protocol.** Starting from an initially randomly loaded 2D atom array, the optimal matching path between the loaded atoms and target

sites is first calculated. The entire path is then divided into *N* steps. For each small step, an AI model is used to calculate the hologram for the SLM and, in real time, the SLM simultaneously moves all atoms to their target positions. After switching to the pre-defined target array, a defect-free atom array is achieved.

**Fig. 2 | Schematic of the experimental set-up.** The optical tweezer array is generated by using a high-speed SLM to imprint an appropriate hologram onto the incident laser beam, then focusing the beam through a 0.55-NA objective. The fluorescence of the atoms trapped in the tweezer array is separated from the tweezer beam by a dichroic mirror (DM) and detected using an EMCCD. An electrically tunable lens (ETL), whose focal length can be changed, allows imaging of different planes of a 3D atom array. Based on the atom occupancy, an AI model calculates the next series of holograms for SLM refresh, which enables parallel atom movement (inset).

**Fig. 3 | AI model. a,** Training and inference processes of the AI model for hologram generation. In the training stage, the coordinates of the tweezer array, with a resolution of 8192, are obtained by simulating each step of the rearrangement process. These coordinates are fed into the WGS algorithm to generate the corresponding holograms (8192×8192). Each hologram is transformed into the amplitude and phase images $\{A_{tweezer}, \phi_{tweezer}\}$ in the position domain using FFT. The phase image $\phi_{tweezer}$ is then used to extract the phase values at the tweezers coordinates. Meanwhile, the central 1024×1024 region of the hologram is also transformed by FFT to serve as the target supervision $\{A_{label}, \phi_{label}\}$. The inputs of the CNN, $A_{input}$ and $\phi_{input}$ (1024×1024), are encoded by interpolating the tweezers coordinates and their

corresponding phase values. The network predicts $A_{output}$ and $\phi_{output}$, which are compared with the target supervision. The differences serve as the training loss, and the model is optimized by minimizing this loss. In practical inference, the AI model receives the real-number tweezers coordinates and phases, and the final hologram (1024×1024) is obtained by applying an IFFT to the predicted amplitude and phase. **b,** Precise control over the positions and phases of optical tweezers by the AI model. The positions and phases of the tweezers are simulated and compared with the input values. (Top) The position displacement is isotropic in all directions, with a standard deviation of about 20 nm. (Bottom) The standard deviation of the phase error is about 0.2 rad.

**Fig. 4 | Gallery of thousands of defect-free 2D and 3D atom arrays. a,** Nearly defect-free 2D square array with 2024 atoms. The target array had 45×45=2025 sites, with one atom lost. **b,** Fully defect-free 2D pattern of the letters 'USTC' with 723 atoms. **c,** Nearly defect-free trilayer cuboid array with 1077 atoms. The target array had 19×19×3=1083 sites, with 6 atoms lost. The 3D array was achieved in a single rearrangement, and the image was obtained by tomographic imaging of different layers. **d,** Nearly defect-free trilayer twisted graphene structure with 752 atoms. The twist angle between adjacent layers was 20 degrees. The target array had 252×3=756 sites, with 4 atoms lost. The main image shows a top-down view, which clearly displays the Moiré pattern, while the inset illustrates the 3D structure.

**Fig. 5 | Rearrangement results. a,** Statistical distribution of the atom number in the target arrays from 1000 repetitions of the experiments. A single round of rearrangement (pink) results in filling fraction of 99.0(2)%, with a maximum recorded atom number

of 2,016. Two rounds of rearrangement (blue) result in filling fraction of 99.6(2)%, with the maximum recorded atom number reaching 2,024. **b,** Simulated time overhead for each subroutine for the atom rearrangement. Path matching (pink), hologram computation on two GPUs (blue), and SLM refresh (purple) costs 5 ms, 52 ms, and 20 ms, respectively. These time cost remains constant for different atom array size from 1,000 to 10,000. The GPU computation and SLM refresh are performed simultaneously, thus the entire rearrangement process takes less than 60 ms.

## **Competing interests**

Some of the technologies described in this article are included in a patent application (Application No. PCT/CN2024/136540), and some of the authors are listed as inventors.

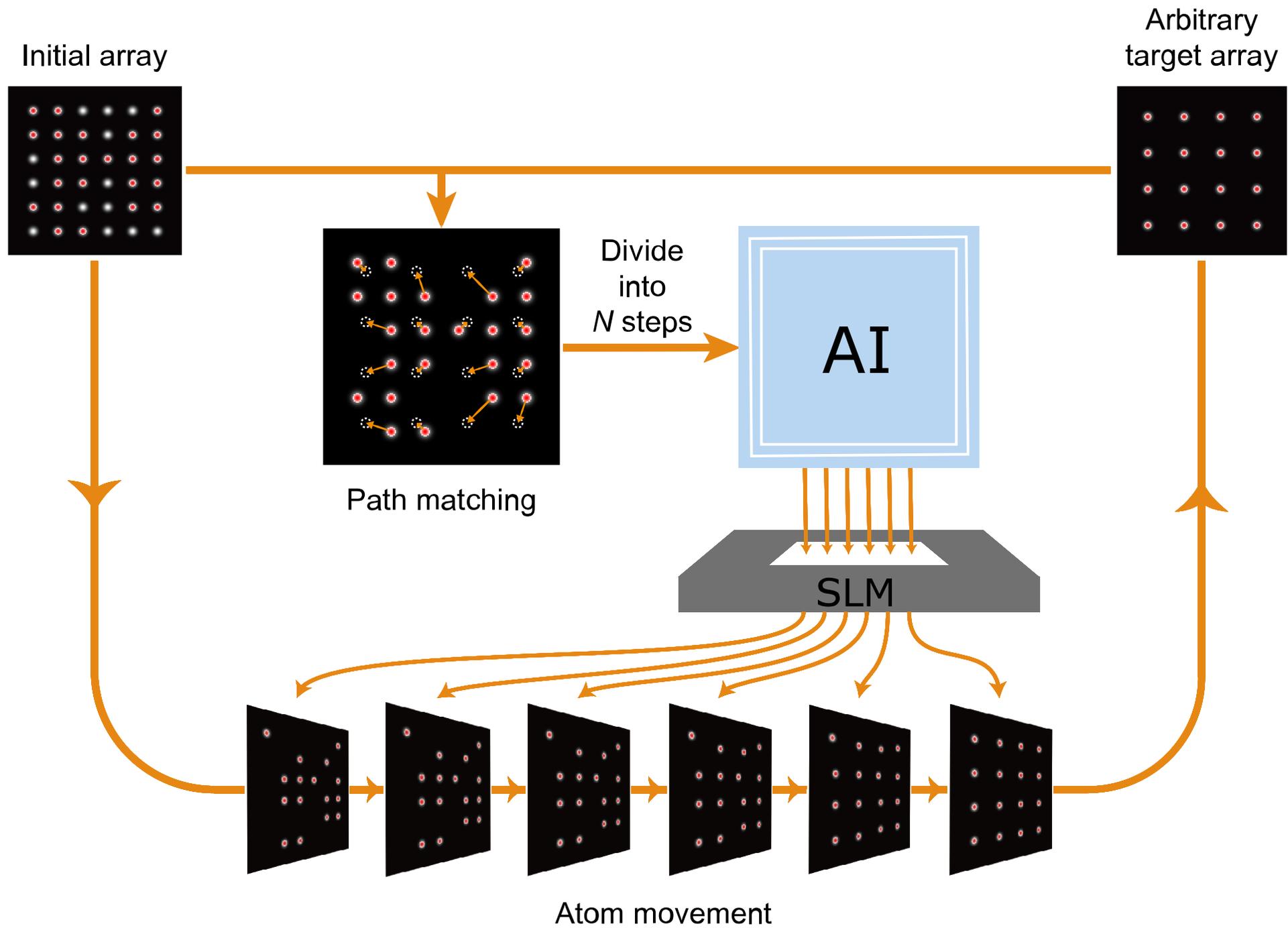

Fig.1

**Fig.2**

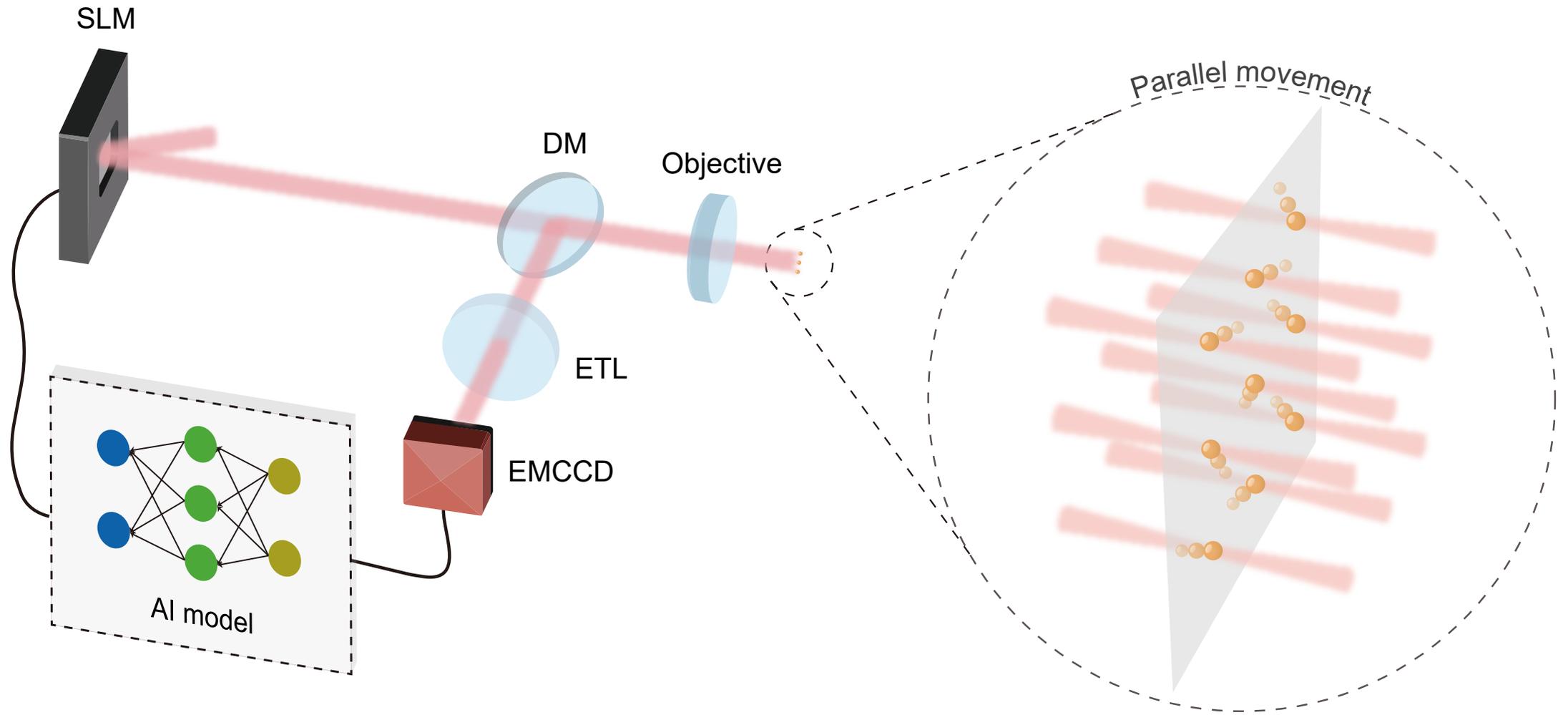

**Fig.3**

a 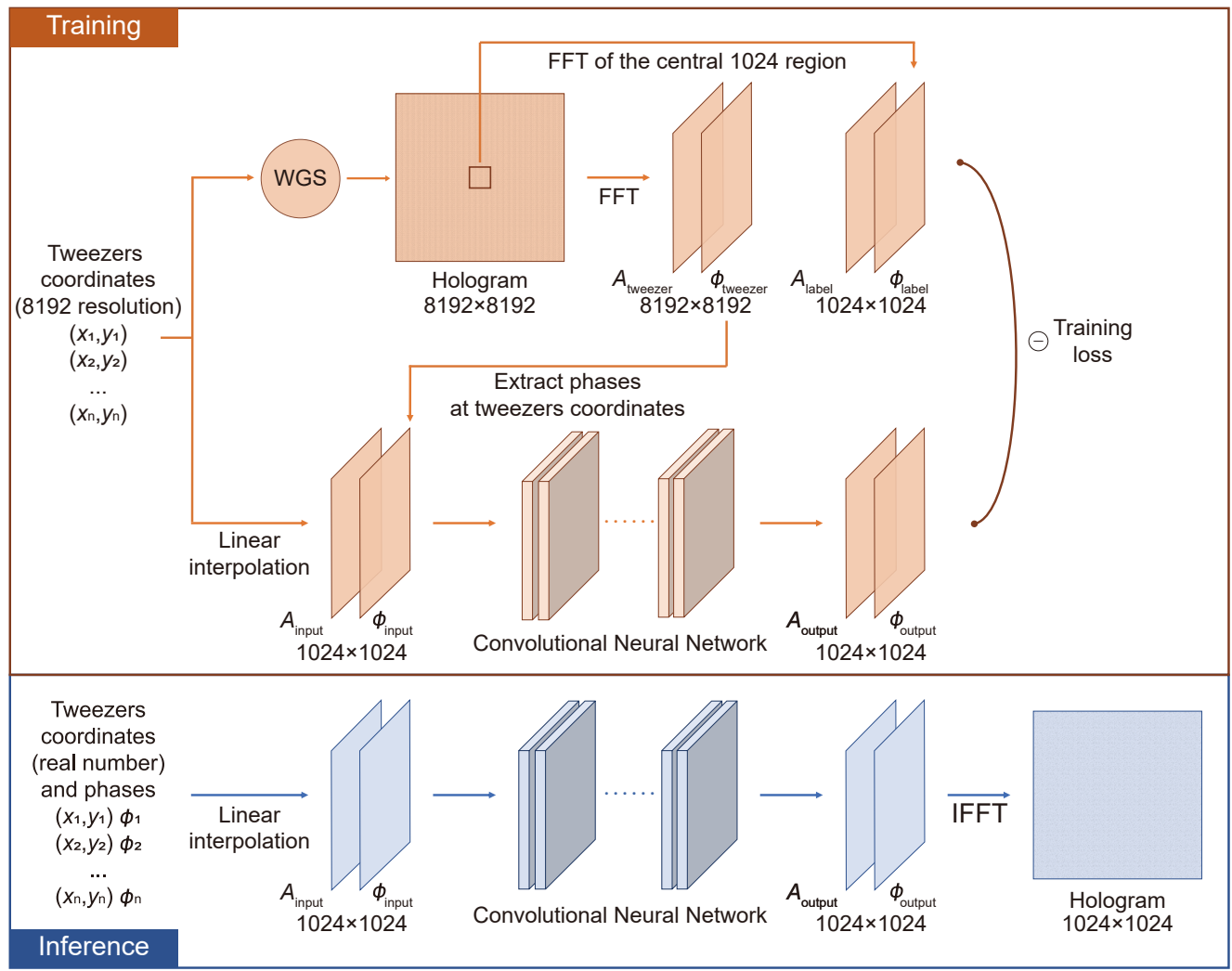

b 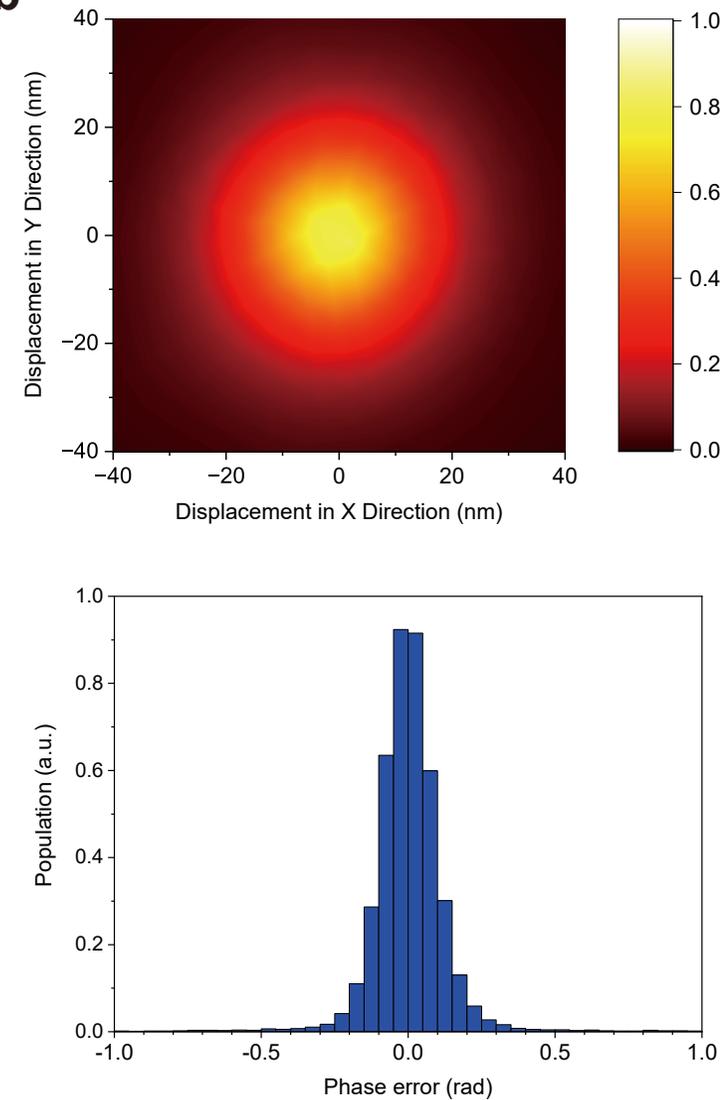

**Fig.4**

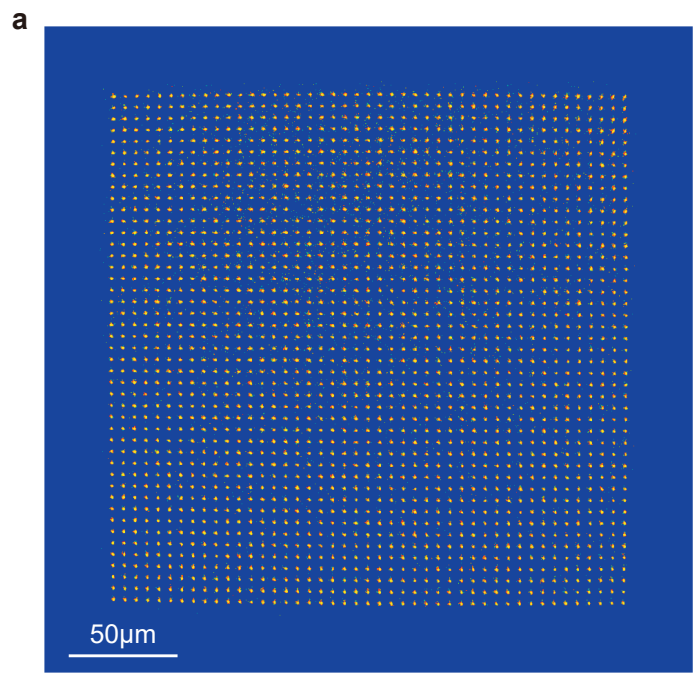 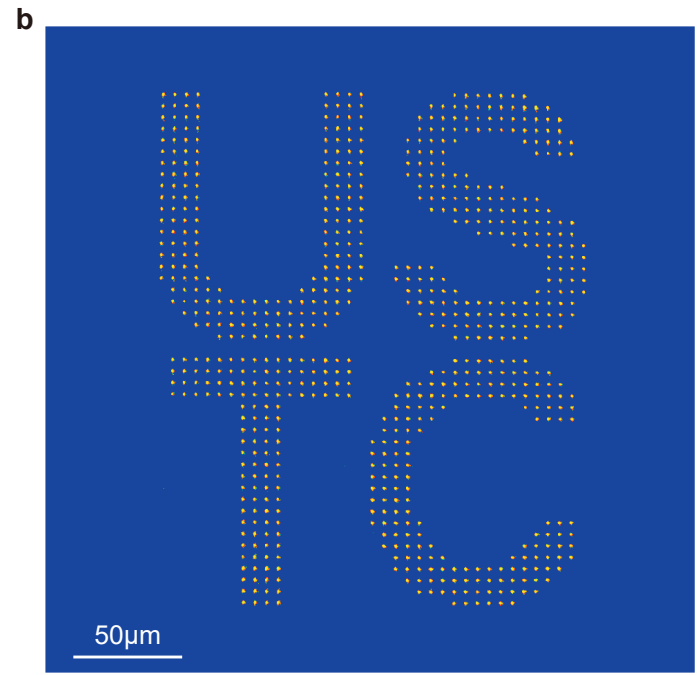
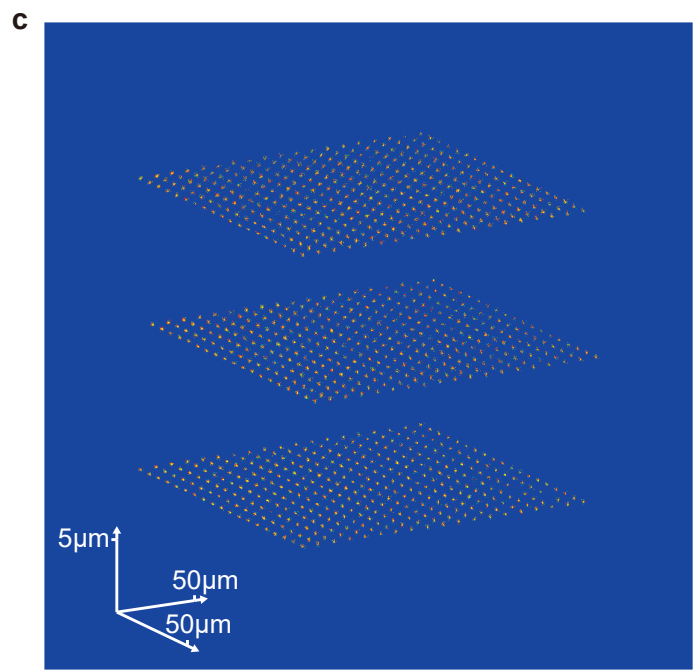 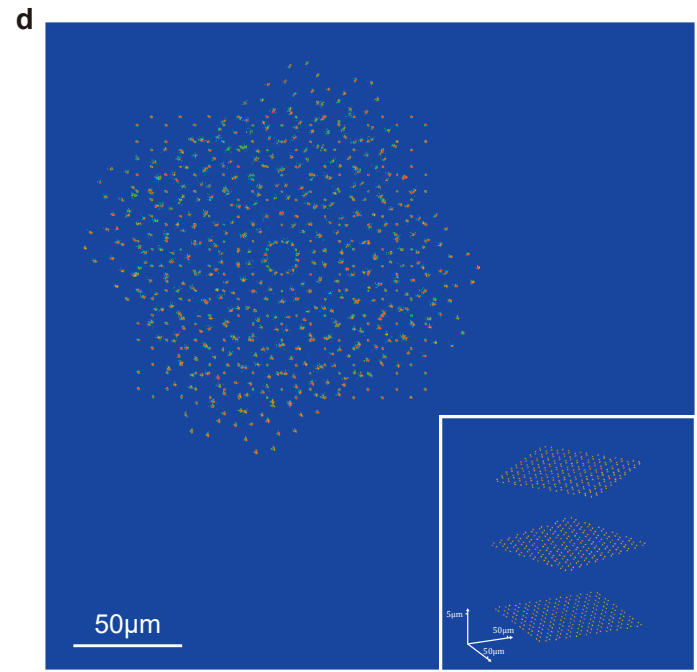

**Fig.5**

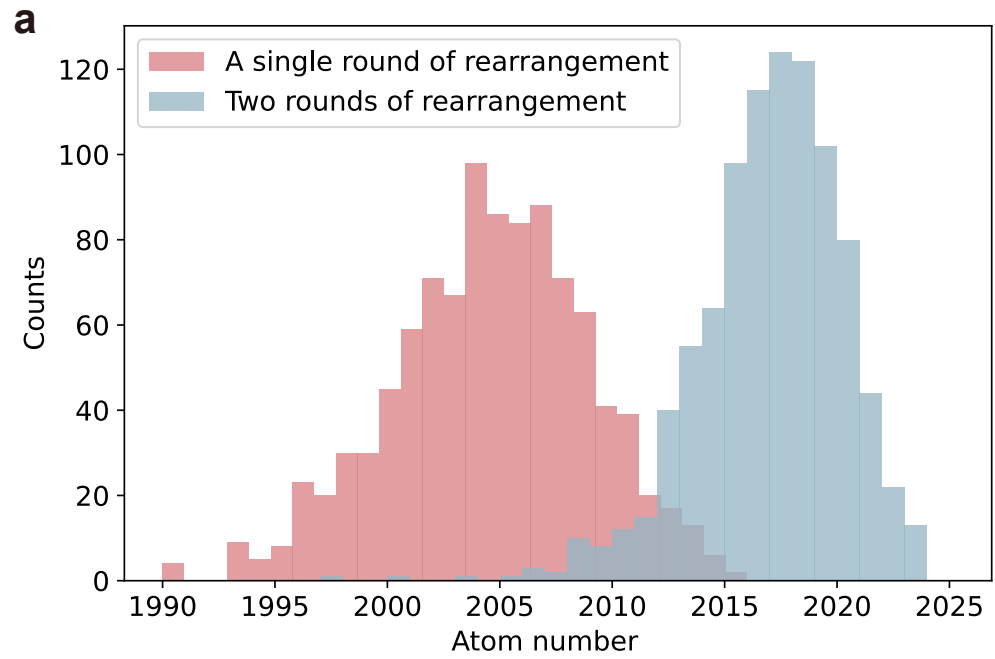

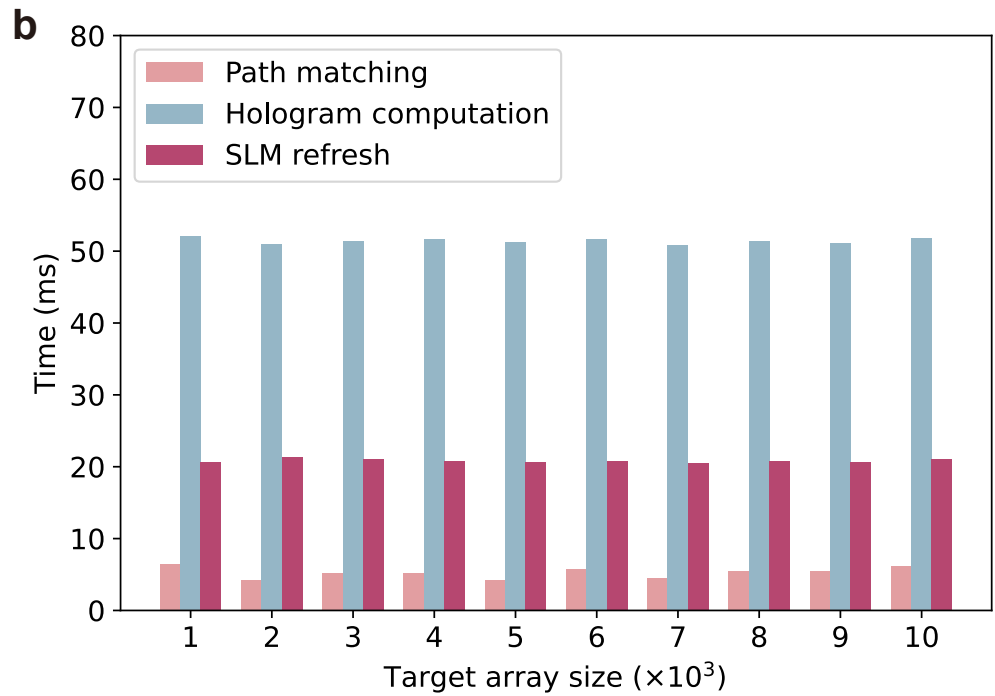